\documentclass[12pt]{article}
\usepackage{amsfonts}
\usepackage{amsmath}
\usepackage{amssymb}
\usepackage{geometry}
\usepackage{apacite}
\usepackage{apalike}
\usepackage{setspace}
\usepackage{float}
\usepackage{portland}
\usepackage{scalefnt}
\usepackage{color}

\input{tcilatex}

\begin{document}

\begin{center}
\begingroup%
\scalefont{1.7}%
\textbf{On Bayesian Nonparametric Continuous Time Series Models}\endgroup%
\medskip

\bigskip \bigskip

\textbf{George Karabatsos\medskip \footnote{%
\noindent Corresponding author. \noindent Professor, University of
Illinois-Chicago, U.S.A., Program in Measurement and Statistics. 1040 W.
Harrison St. (MC\ 147), Chicago, IL 60607. E-mail: gkarabatsos1@gmail.com.
Phone:\ 312-413-1816.}}\\[0pt]
\textit{University of Illinois-Chicago}

\bigskip

\noindent and

\bigskip

\textbf{Stephen G. Walker\medskip \footnote{%
\noindent Professor, University of Kent, United Kingdom, School of
Mathematics, Statistics \& Actuarial Science. Currently, Visiting Professor
at The University of Texas at Austin, Division of Statistics and Scientific
Computation.}}

\textit{University of Kent, United Kingdom}\\[0pt]

\bigskip

March 2, 2013
\end{center}

\bigskip

\noindent \textbf{Abstract:} This paper is a note on the use of Bayesian
nonparametric mixture models for continuous time series. We identify a key
requirement for such models, and then establish that there is a single type
of model which meets this requirement. As it turns out, the model is well
known in multiple change-point problems.


\newpage

\section{Introduction{\textbf{\label{Introduction Section}}}}

A lot of recent research has focused on the development of Bayesian
nonparametric, countably-infinite mixture models for time series data. This
work has aimed to relax the normality assumptions of the general class of
dynamic linear models (West \&\ Harrison, 1997\nocite{WestHarrison97}),
which already encompasses traditional normal (time-static)\ linear
regression, autoregressions, autoregressive moving average (ARMA) models,
and nonstationary polynomial trend and time-series models.

These Bayesian nonparametric (infinite-mixture)\ time series models have the
general form:%
\begin{equation*}
f(y_{t}|t)=\dint f(y_{t}|\mathbf{x}_{t},\boldsymbol{\gamma },\boldsymbol{%
\theta })\mathrm{d}G_{t}(\boldsymbol{\theta })=\tsum\limits_{j=1}^{\infty
}f(y_{t}|\mathbf{x}_{t},\boldsymbol{\gamma },\boldsymbol{\theta }%
_{tj})\omega _{j}(t),
\end{equation*}%
given time $t\in \mathcal{T};$ kernel (component) densities $\{f(\cdot |%
\mathbf{x}_{t},\boldsymbol{\gamma },\boldsymbol{\theta }_{tj})\}_{j=1}^{%
\infty }$ which are often specified by normal densities of a dynamic linear
model; mixture distribution%
\begin{equation*}
G_{t}(\cdot )=\tsum\nolimits_{j=1}^{\infty }\omega _{j}(t)\delta _{%
\boldsymbol{\theta }_{j}(t)}(\cdot ),
\end{equation*}%
which is formed by an infinite-mixture of point-mass distributions $\delta _{%
\boldsymbol{\theta }_{j}(t)}(\cdot )$ with mixture weights $\omega _{j}(t)$;
and prior distributions $\boldsymbol{\gamma }\sim \pi (\boldsymbol{\gamma })$%
,$\ \boldsymbol{\theta }_{tj}\sim G_{0t}$, $\{\omega _{j}(t)\}_{j=1}^{\infty
}\sim \Pi $. \ All of the earlier models focuses on discrete time, and
specify $G_{t}$ to be some variant of the Dependent Dirichlet process (DDP)\
(MacEachern, 1999\nocite{MacEachern99}, 2000\nocite{MacEachern00}, 2001%
\nocite{MacEachern01}), so that the mixture weights have a stick-breaking
form, with 
\begin{equation*}
\omega _{j}(t)=\upsilon _{j}(t)\tprod\nolimits_{l=1}^{j=1}(1-\upsilon
_{l}(t)),\text{ and }\upsilon _{j}(t):\mathcal{T}\rightarrow \lbrack 0,1].
\end{equation*}%
(Sethuraman, 1994\nocite{Sethuraman94}). Such DDP-based time-series models
either assume time-dependent stick-breaking weights (Griffin \&\ Steel, 2006%
\nocite{GriffinSteel06}, 2011; Rodriguez \&\ Dunson, 2011\nocite%
{RodriguezDunson11}), or assume non-time-dependent stick-breaking weights
and a time-dependent prior (baseline) distribution $G_{0t}$ (Rodriguez \&\
ter Horst, 2008\nocite{RodriguezTerHorst08}), or assume a fully
non-time-dependent Dirichlet process (DP)\ $G_{t}=G$ with only
time-dependence in the kernel densities (Hatjispyros, et al., 2009\nocite%
{Hatjispyros_Nicoleris_Walker09}; Tang \& Ghosal, 2007\nocite{TangGhosal07};
Lau \&\ So, 2008\nocite{LauSo08}; Caron et al., 2008\nocite%
{CaronDavyDoucetDuflos08}; Giardina et al., 2011\nocite{Giardina_etal11}; Di
Lucca et al., 2012\nocite{DiLucca_etal12}). Other related approaches
construct a time-dependent DDP$\ G_{t}$ either by generalizing the P\'{o}lya
urn scheme of the DP (e.g., Zhu et al., 2005\nocite{Zhu_etal05}; Caron et
al., 2007\nocite{CaronDavyDoucet07}); by a convex combination of
hierarchical Dirichlet processes (HDP)\ or DPs (Ren et al., 2008\nocite%
{Ren_etal08}; Dunson, 2006\nocite{Dunson06}); by a HDP-based hidden Markov
model that has infinitely-many states (Fox et al., 2008\nocite{Fox_etal08},
2011\nocite{Fox_etal11}); or by a Markov-switching model having
finitely-many states (Taddy \& Kottas, 2009\nocite{TaddyKottas09}).

The more recent work on Bayesian nonparametric time series modeling has
focused on continuous time, and on developing a time-dependent mixture
distribution that has the general form, 
\begin{equation}
G_{t}=\dsum\limits_{j=1}^{\infty }\omega _{j}(t)\delta _{\boldsymbol{\theta }%
_{j}}(\cdot ),  \label{GenCont}
\end{equation}%
based on a process other than the DDP. Above, a baseline prior $\boldsymbol{%
\theta }_{j}\sim _{iid}$ $G_{0}$ is assumed, which is a standard assumption.

In Section 2 we describe these continuous time series models, namely, the
geometric model (Mena, Ruggiero, \&\ Walker, 2011\nocite%
{MenaRuggieroWalker11}), and a normalized random measure model (NRM)\
(Griffin, 2011\nocite{Griffin11}). In Section 3, we highlight a key property
such models are required to possess, and we identify a necessary model which
has such a property. We also in this section prove that the existing
continuous time series models do not have the required property.

\section{Continuous time models{\textbf{\label{CTM Section}}}}

The geometric model constructs a dependent process $G_{t}$ using
time-dependent geometric mixture weights%
\begin{equation}
\omega _{j}(t)=\lambda _{t}(1-\lambda _{t})^{j-1},  \label{GEOmix}
\end{equation}%
with $\lambda _{t}$ specified as a two-type Wright-Fisher diffusion (Mena,
Ruggiero, \&\ Walker, 2011\nocite{MenaRuggieroWalker11}).

The $(\lambda )_{t}$ follow a stochastic process with the stationary density
being a beta$(a,b)$. The transition mechanism is given, for $t>s$, by 
\begin{equation*}
p(\lambda _{t}|\lambda _{s})=\sum_{m=0}^{\infty }p_{h}(m)\,p(\lambda
_{t}|m,\lambda _{s})
\end{equation*}%
where $h=t-s$ and 
\begin{equation*}
p(\lambda _{t}|m,\lambda _{s})=\sum_{k=0}^{m}\mathrm{beta}(\lambda
_{t}|a+k,b+m-k)\,\mathrm{bin}(k|m,\lambda _{s})
\end{equation*}%
and 
\begin{equation*}
p_{h}(m)=\frac{(a+b)_{m}\,\exp (-mch)}{m!}\,\left( 1-e^{-ch}\right) ^{a+b}
\end{equation*}%
for some $c>0$.

Hence $G_{t}$ is a continuous time process and the properties are studied in
Mena, Ruggiero, and\ Walker (2011\nocite{MenaRuggieroWalker11}).

The normalized random measures (NRM)\ model constructs a time-dependent
process $G_{t}$ using time-dependent mixture weights that are formed by
normalizing a stochastic process derived from non-Gaussian
Ornstein-Uhlenbeck processes (Griffin, 2011\nocite{Griffin11}).
Specifically, these weights are constructed by%
\begin{equation}
\omega _{j}(t)=\dfrac{\mathbf{1}(\tau _{j}\leq t)\exp (-\lambda (t-\tau
_{j}))J_{j}}{\tsum\nolimits_{l=1}^{\infty }\mathbf{1}(\tau _{l}\leq t)\exp
(-\lambda (t-\tau _{l}))J_{l}},  \label{NRMmix}
\end{equation}%
where $(\tau ,J)$ follows a Poisson process with intensity $\lambda \,w(J)$,
where $w$ is a L\'{e}vy density.

Details and examples of obtaining the $(\tau _{j},J_{j})$ are provided by
Griffin (2011\nocite{Griffin11}). Aside from the specific examples
considered in this paper, we also note that any sequence of $(\tau
_{j},J_{j})$ are permissible provided 
\begin{equation*}
{\tsum\nolimits_{l=1}^{\infty }\mathbf{1}(\tau _{l}\leq t)\exp (-\lambda
(t-\tau _{l}))J_{l}}<\infty
\end{equation*}%
for all $t$.

\section{A key property{\textbf{\label{AKP Section}}}}

Using the mixture model 
\begin{equation*}
f(y|t)=\int K(y|\theta)\,G_t(d\theta)=\sum_{j=1}^\infty
w_j(t)\,K(y|\theta_j),
\end{equation*}
we insist on the obvious requirement that for all suitably small $h$, we
want $y_t$ and $y_{t+h}$ to be arising from the same component. This
requirement is clearly not met by simply insisting that $G_{t+h}\rightarrow
G_t$ as $h\rightarrow 0$.

So, in this paper, we introduce the argument that a Bayesian nonparametric
continuous time series model should have a certain property. Specifically,
based on the above discussion, we need the property that 
\begin{equation*}
P(\theta _{t}=\theta _{t+h})\rightarrow 1\quad \text{as}\quad h\rightarrow 0,
\end{equation*}%
where $\theta _{t}$ denotes a sample from 
\begin{equation*}
G_{t}=\sum_{j=1}^{\infty }\omega _{j}(t)\,\delta _{\theta _{j}},
\end{equation*}%
i.e. that $\theta_t|G_t\sim G_t$, which means that $P(\theta_t=%
\theta_j)=w_j(t)$.

Now it can be shown that 
\begin{equation*}
P(\theta _{t}=\theta _{t+h})=\sum_{j=1}^{\infty }P(\theta _{t}=\theta
_{t+h}=\theta _{j}),
\end{equation*}%
and hence we are asking for 
\begin{equation*}
\text{\textrm{E}}\left\{ \sum_{j=1}^{\infty }\omega _{j}(t)\,\omega
_{j}(t+h)\right\} \rightarrow 1\quad \text{as}\quad h\rightarrow 0.
\end{equation*}%
For this, it is necessary that 
\begin{equation*}
D(h)=\sum_{j=1}^{\infty }\omega _{j}(t)\,\omega _{j}(t+h)\rightarrow 1\quad 
\text{in probability as}\quad h\rightarrow 0.
\end{equation*}%
Now assume that 
\begin{equation*}
\sup_{j}|\omega _{j}(t+h)-\omega _{j}(t)|\rightarrow 0\quad \mbox{a.s.   as}%
\quad h\rightarrow 0
\end{equation*}%
which is an extremely mild condition.

Hence, for any $\epsilon >0$, 
\begin{equation*}
\sup_{j}|\omega _{j}(t+h)-\omega _{j}(t)|<\epsilon
\end{equation*}%
for all small enough $h$. Therefore, for all small $h$, we have 
\begin{equation*}
D(h)\leq \sum_{j=1}^{\infty }\omega _{j}^{2}(t)+\epsilon \quad \mbox{a.s.}
\end{equation*}%
The only way we can now recover the convergence to 1 in probability is that 
\begin{equation*}
\omega _{j}(t)=1\quad\mbox{a.s.}
\end{equation*}%
for a particular $j$, which will depend on $t$.

Hence, we believe that a Bayesian nonparametric continuous time series model
should specify a time-dependent mixture distribution $G_{t}$ of the type
given in (\ref{GenCont}), where 
\begin{equation*}
\omega _{j}(t)=\mathbf{1}(t\in A_{j}),
\end{equation*}%
and the $(A_{j})_{j}$ form a random partition of $(0,\infty )$. In other
words, we recommend Bayesian nonparametric change-point mode for time series
analysis. Specifically, let $\mathcal{D}=\{(y_{t_{i}})\}_{i=1}^{n}$ denote a
sample of data consisting of $n$ dependent responses $y_{t_{i}}$ observed at
time points $t_{i}$. Then, such a model may be specified as: 
\begin{subequations}
\label{NewModelChange}
\begin{eqnarray}
y_{t_{i}} &\sim &f(y_{t_{i}}|\boldsymbol{\theta }_{z(t_{i})}),\text{ }%
i=1,\ldots ,n, \\
z(t_{i}) &=&j\iff \tau _{j-1}<t_{i}\leq \tau _{j}=(\tau _{j-1}+\epsilon _{j})
\\
\epsilon _{j} &\sim &\text{ }\mathrm{Ex}(\lambda ),\text{ \ }j=1,2,\ldots 
\text{,} \\
\boldsymbol{\theta }_{j} &\sim &G_{0},\text{ \ }j=1,2,\ldots \text{,}
\end{eqnarray}%
where $z(t_{i})$ denotes the random component index, and each of the gaps $%
\epsilon _{j}=\tau _{j}-\tau _{j-1}$ are i.i.d. from an exponential $\mathrm{%
Ex}(\lambda )$ prior distribution, with $\tau _{0}:=0$. The exponential
distribution for creating the intervals is not essential but there seems
little reason to make it more complicated.

Interestingly, neither the geometric model nor the NRM model specify a
mixing distribution $G_{t}$ with weights that satisfy the key property,
previously described. Figure \ref{GeometricPlot} illustrates this fact.
Specifically, for the geometric model, the figure shows samples of the
random component index $z(t)\sim \Pr (z(t)=j)\propto \omega _{j}(t)$, over a
convergent sequence of times $t=t_{l-1}+1/l^{2},$ for $l=1,2,\ldots ,1000,$
with $t_{0}=0$. These samples are presented for different choices of prior
parameters in this model, namely $b=1,10,30,50,$ along with $a=c=1$. As the
figure shows, as $t$ converges to time $1.6439,$ the random variable $z(t)$
does not converge to a single value. Instead, the random variable displays a
degree of uncertainty about the component (kernel)\ density at that time.

Now, we formally show how our time series model satisfies the property,
whereas the geometric model and the NRM models do not. For our model for
which we have based on the 
\end{subequations}
\begin{equation*}
w_{j}(t)=\mathbf{1}(t\in A_{j})
\end{equation*}%
and 
\begin{equation*}
A_{j}=(\tau _{j-1},\tau _{j})
\end{equation*}%
with 
\begin{equation*}
\tau _{j}=\tau _{j-1}+\epsilon _{j}
\end{equation*}%
where the $(\epsilon _{j})$ are independent and identically distributed
exponential random variables with parameter $\lambda $, it is
straightforward to show that 
\begin{equation*}
\sum_{j=1}^{\infty }w_{j}(t)\,w_{j}(t+h)=\left\{ 
\begin{array}{ccc}
1 & \text{with probability} & e^{-h} \\ 
0 & \text{with probability} & 1-e^{-h}.%
\end{array}%
\right.
\end{equation*}%
This follows since we need $t,t+h\in A_{j}$. Hence, it it seen that 
\begin{equation*}
\mathrm{E}\left\{ \sum_{j=1}^{\infty }w_{j}(t)\,w_{j}(t+h)\right\}
=e^{-h}\rightarrow 1\quad \text{as}\quad h\rightarrow 0.
\end{equation*}

For the geometric model, we have 
\begin{equation*}
\mathrm{E}\left\{ \sum_{j=1}^{\infty }w_{j}(t)\,w_{j}(t+h)\right\}
\end{equation*}%
given by 
\begin{equation*}
\mathrm{E}\left\{ \sum_{j=1}^{\infty }\lambda _{t}(1-\lambda
_{t})^{j-1}\,\lambda _{t+h}(1-\lambda _{t+h})^{j-1}\right\}
\end{equation*}%
which is 
\begin{equation*}
\mathrm{E}\left\{ \frac{\lambda _{t}\lambda _{t+h}}{\lambda _{t}+\lambda
_{t+h}-\lambda _{t}\lambda _{t+h}}\right\} .
\end{equation*}%
This is strictly less than one due to the fact that $\lambda _{t}$ and $%
\lambda _{t+h}$ are less than 1.

Finally, the NRM model also has%
\begin{equation*}
\mathrm{E}\left\{ \sum_{j=1}^{\infty }w_{j}(t)\,w_{j}(t+h)\right\} <1,
\end{equation*}%
and this result follows from the proof of his Theorem 2, which appears in
the Appendix of his paper.

\section{Discussion{\textbf{\label{Discussion Section}}}}

In summary, we advocate a specific property for mixture models for
continuous time series. Namely, that as the time $t+h$ approaches the limit $%
h\rightarrow 0$, the model should certainly identify a single component
index $z(t),$ and hence a single component density $f(y_{t}|\boldsymbol{%
\theta }_{z(t)})$ of the dependent response $Y_{t}$. In other words, there
is no strong reason why one should specify a time-series model that allows
the component density to drastically change, as time goes through
incrementally smaller changes. In essence we are not asking for $G_{t}$ to
be close to $G_{t+h}$, though this is given, but a rather weak condition;
rather we are asking that $\theta _{t}$ and $\theta _{t+h}$ are close in
probability, which approaches 1 as $h\rightarrow 0$.

Interestingly, we have shown that two Bayesian nonparametric
(infinite-mixture) models fail this sensible property. In contrast, we have
shown that for a mixture model to satisfy the property, it must be of the
form given in equation (\ref{NewModelChange}). This implies that the mixture
model must be a Bayesian multiple change-point model (e.g., Barry \&
Hartigan, 1993\nocite{BarryHartigan93}; Chib, 1998\nocite{Chib98}), having
infinitely-many change-point parameters $\tau _{j-1}<\tau _{j},$ $%
j=1,2,\ldots $. Then, these results may encourage future developments in
Bayesian nonparametric models for continuous time series, more in terms of
multiple change point modeling.

\section{\noindent \noindent Acknowledgements\ }

This research is supported by National Science Foundation Grant SES-1156372,
from the Program in Methodology, Measurement, and Statistics.

\bibliographystyle{apacite}
\bibliography{Karabatsos}

\newpage \FRAME{fthFU}{6.5965in}{3.0752in}{0pt}{\Qcb{For the geometric time
series model, the log of samples of component index $z(t)\sim \Pr
(z(t)=j)\propto \protect\omega _{j}(t)$, over a convergent sequence of times 
$t=t_{l-1}+1/l^{2},$ for $l=1,2,\ldots ,1000,$ with $t_{0}=0$. The component
index samples are shown for a range of choices of prior parameters, $%
b=1,10,30,50$, along with $a=c=1$.}}{\Qlb{GeometricPlot}}{timemodels.eps}{%
\special{language "Scientific Word";type "GRAPHIC";maintain-aspect-ratio
TRUE;display "USEDEF";valid_file "F";width 6.5965in;height 3.0752in;depth
0pt;original-width 5.8208in;original-height 2.7003in;cropleft "0";croptop
"1";cropright "1";cropbottom "0";filename '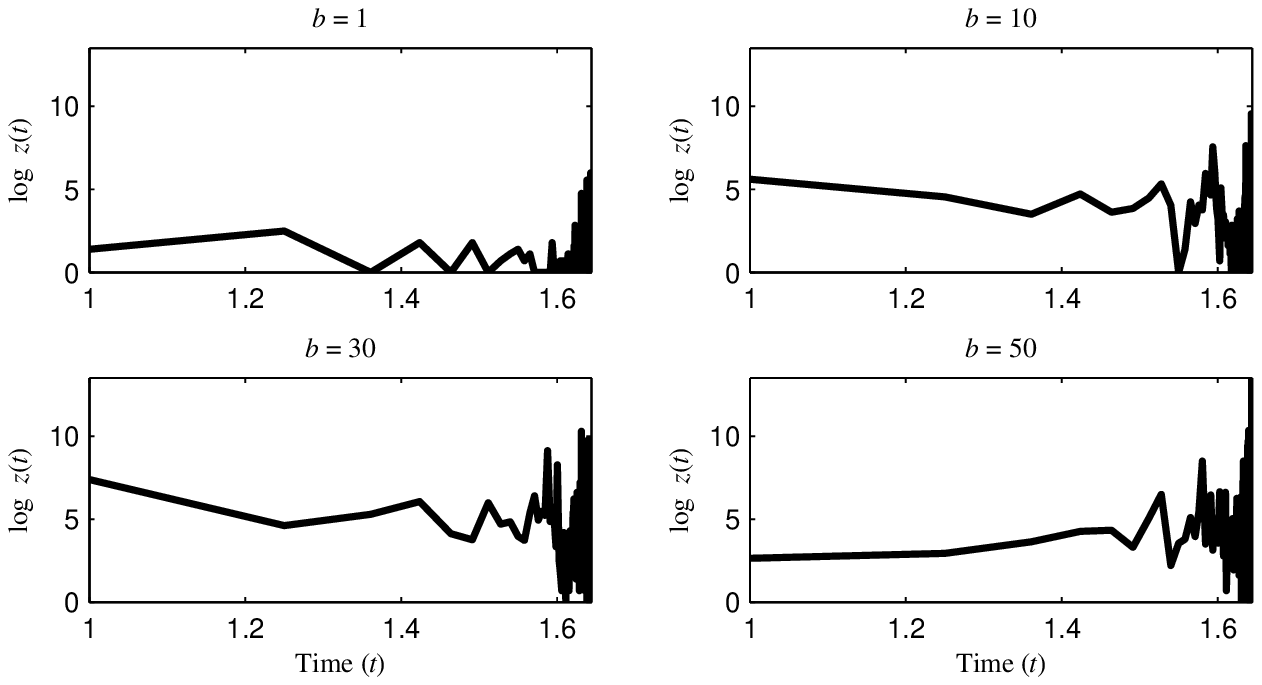';file-properties
"XNPEU";}}

\end{document}